\documentclass[prb, reprint, 
  superscriptaddress]{revtex4-1}
\usepackage[utf8]{inputenc}
\usepackage{graphicx}
\usepackage{amssymb}
\usepackage{amsmath}
\usepackage{color}
\usepackage{hyperref}

\begin{document}

\title{Critical behavior of the damping rate of GHz acoustic phonons in SrTiO$_3$ at the antiferrodistortive phase transition measured by time- and frequency-resolved Brillouin scattering}

\author{Lena Maerten}
\affiliation{Institut f\"ur Physik \& Astronomie, Universit\"at Potsdam, Karl-Liebknecht-Str. 24-25, 14476 Potsdam, Germany}

\author{Andr\'e Bojahr}
\affiliation{Institut f\"ur Physik \& Astronomie, Universit\"at Potsdam, Karl-Liebknecht-Str. 24-25, 14476 Potsdam, Germany}

\author{Matthias Reinhardt}
\affiliation{Helmholtz-Zentrum-Berlin f\"ur Materialien und Energie, Wilhelm-Conrad-R\"ontgen Campus, BESSY II, Albert-Einstein-Str. 15, 12489 Berlin, Germany}

\author{Akitoshi Koreeda}
\affiliation{Department of Physical Sciences, College of Science and Engineering, Ritsumeikan University, 1-1-1 Noji-Higashi, Kusatsu, Shiga pref., 525-8577 Japan}

\author{Matthias R\"ossle}
\email{matthias.roessle@helmholtz-berlin.de}
\affiliation{Institut f\"ur Physik \& Astronomie, Universit\"at Potsdam, Karl-Liebknecht-Str. 24-25, 14476 Potsdam, Germany}
\affiliation{Helmholtz-Zentrum-Berlin f\"ur Materialien und Energie, Wilhelm-Conrad-R\"ontgen Campus, BESSY II, Albert-Einstein-Str. 15, 12489 Berlin, Germany}

\author{Matias Bargheer}
\email{bargheer@uni-potsdam.de}
\homepage{www.udkm.physik.uni-potsdam.de}
\affiliation{Institut f\"ur Physik \& Astronomie, Universit\"at Potsdam, Karl-Liebknecht-Str. 24-25, 14476 Potsdam, Germany}
\affiliation{Helmholtz-Zentrum-Berlin f\"ur Materialien und Energie, Wilhelm-Conrad-R\"ontgen Campus, BESSY II, Albert-Einstein-Str. 15, 12489 Berlin, Germany}

\date{\today}

\begin{abstract}
We determine the temperature dependent damping rate of longitudinal acoustic phonons in SrTiO$_3$ using frequency domain Brillouin scattering and time domain Brillouin scattering. We investigate samples with (La,Sr)MnO$_3$ and SrRuO$_3$ capping layers, which result in compressive or tensile strain at the layer -- substrate interface, respectively. The different strain states lead to different domain structures in SrTiO$_3$ that extend into the bulk of the SrTiO$_3$ substrates and strongly affect the phonon propagation. Our experiments show that the damping rate of acoustic phonons in the interfacial STO layer depends strongly on the sample temperature and strain induced domain structure. We also show that the damping rate as function of temperature exhibits a critical behavior close to the cubic-to-tetragonal phase transition of SrTiO$_3$.  
\end{abstract}

\maketitle

\section{Introduction}
\label{sec:introduction}

SrTiO$_3$ (STO) is one of the prototypical perovskite oxide materials. It attracted a lot of attention due to its incipient ferroelectric properties where quantum fluctuations inhibit the formation of a long-range ferroelectric order even at lowest temperatures~\cite{Muller1979}. It is also technologically important as its large dielectric constant~\cite{Muller1979} as well as its bulk lattice constant~\cite{Lytle1964} of $a\!=\!0.3905$\,nm allows to tailor device structures with new artificial properties comprising of other, especially perovskite based, materials. It is for example possible to epitaxially grow thin high-temperature superconducter films or other materials with magnetic or ferroic properties on STO substrates and study interactions within these heterostructures.

STO crystallizes the so-called perovskite crystal structure. At room temperature ($T$), an oxygen octahedron surrounds the central Ti cation. The octahedron is embedded in a cubic arrangement of Sr cations. STO undergoes a structural, antiferrodistortive (AFD) phase transition at $T_{\text{a}}\!\approx\!105$\,K where the octahedra of adjacent unit cells rotate along a common axis in opposite directions~\cite{Shirane1969}. This phase transition results in a slight tetragonal distortion of the unit cell below $T_{\text{a}}$ with $^c\!/\!_a\approx1.0006$. Below $T_{\text{a}}$, the tetragonal axes are in individual domains aligned parallel to a common rotation axis~\cite{Lytle1964}. The orientation of the $c$ axes depends on local strain fields due to crystal imperfections and occurs with almost equal probability along any of three cartesian coordinate axes~\cite{Sidoruk2010}. It is well-known that the AFD transition takes place at the $R$ point of the Brillouin zone boundary and involves the softening of the corresponding triply degenerate phonon mode~\cite{Cowley1996a,Carpenter2007,Perks2014}. This structural transition produces domain patterns with needle-shaped twin domains~\cite{Lytle1964,Buckley1999,Salje2012,Honig2013,Kalisky2013}. It is commonly accepted that the domain structure and its stability is important for the mechanical properties of STO below $T_{\text{a}}$~\cite{Bell1963,Kaiser1966,Nava1969,Berre1969,Tagantsev2001}. One for example finds a drop of the sound velocity $v_{\text L}$ of longitudinal acoustic (LA) phonons that for low frequencies on the order of several Hz has been attributed to the so-called superelastic effect~\cite{Kityk2000,Schranz2011}. However, it has recently been shown that also GHz phonons induce the superelastic softening effect in STO because the compressive part of the acoustic phonon mode overcompensates the tetragonal distortion along the phonon propagation direction and thus enhances the mobility of twin domain walls~\cite{Maerten2015}. It has furthermore been proposed that the polarization reversal of ferroelectric domains is accompanied by a supersonic movement of kinks at GHz frequencies~\cite{Salje2017}. In addition, externally applied pressure or electric fields also affect the formation of the domain structure~\cite{Sidoruk2010}. In the vicinity of the surface of bulk STO crystals for example a distorted layer is formed~\cite{Rutt1997,Salman2011} that depends on the sample fabrication~\cite{Hunnefeld2002} and can be modified by strain exerted by thin films epitaxially grown on top~\cite{Loetzsch2010}.

As the alignment of the domains modifies the mechanical properties of STO, it is therefore important to relate the structural to the mechanical properties. A way to study the mechanical properties of materials is to use LA phonons and investigate their propagation properties, that is, their propagation velocity and damping rate. The phonon propagation velocity, $v_{\text L}$, and the elastic moduli along the propagation direction of the phonon are given by the components $c_{ij}$ of the stiffness tensor $\mathbf{C}_{ij}$, which allows to relate both~\cite{Berre1969,Rehwald1970}: For example, the sound velocity $v_{\text L}$ of LA phonons propagating along the $[100]$ direction is given by $v_{\text L}\!=\!\sqrt{^{c_{11}}\!/\!_{\varrho}}$ where $\varrho$ is the mass density~\cite{Newnham2005}. It has experimentally been observed that the attenuation of MHz acoustic phonons in STO close to $T_{\text a}$ drastically increases~\cite{Berre1969,Fossheim1972,Fossum1984} and $v_{\text L}$ exhibits a dip~\cite{Kaiser1966}. This might indicate an increase of the interaction between optical and acoustic phonons~\cite{Berre1969,Rehwald1971}, which in turn gives rise to critical scattering observed in neutron~\cite{Shirane1969,Riste1971} and x ray scattering experiments~\cite{Darlington1975} as well as light scattering~\cite{Kaiser1966,Fleury1968c}. These observations have been embedded in the concept of the so-called critical phenomena where fluctuation amplitudes of the order parameter close to $T_{\text a}$ become so large that the coherence length diverges. As a consequence, the physical description of the phase transition becomes independent of the actual nature of the transition, which is reflected in the term ``universitality class'' that describes the phase transition close to the critical $T$ by a characteristic ``critical'' exponent. For STO, already M\"uller \textit{et al.} have suggested that the order parameter deduced from EPR measurements, the turn angle of the oxygen octahedra, exhibits a critical exponent~\cite{Muller1971} and it has been shown by Berre \textit{et al.} that also the phonon attenuation of MHz phonons can be described assuming a critical behaviour~\cite{Berre1969}.

In this publication we show that the structural modifications of STO beneath the epitaxial metallic nanolayers SrRuO$_3$ and (La$_{0.7}$Sr$_{0.3}$)MnO$_3$ strongly influence the damping rate of GHz LA phonons. Reciprocal space mapping confirms that both materials result in different domain structures at the interface between metallic overlayer and STO substrate. In ``time-resolved'' time-domain Brillouin scattering (TDBS) and ``static'' Brillouin scattering (BS) experiments without time resolution we distinguish between harmonic and anharmonic lattice contributions and determine the damping rate of acoustic phonons as function of $T$. Finally, we show that the damping rate, $\Gamma$, of the LA phonons exhibits a critical behavior with different critical exponents for the samples under compressive and tensile strain above and below $T_{\text a}$, respectively. We report measurements on three different samples: TDBS and x ray diffraction measurements have been performed on two (100)-oriented STO samples (CrysTec, Berlin, Germany; miscut angle 0.1$^\circ$), one with a 37\,nm thick (La$_{0.7}$Sr$_{0.3}$)MnO$_3$ transducer layer (in the following referred to as ``LSMO'') and another one with a 15\,nm thick SrRuO$_3$ transducer layer (``SRO''), both grown by pulsed laser deposition. The lattice mismatch between STO and the transducer materials results in negative (in the case of LSMO) and positive (in the case of SRO) in-plane strain. BS measurements were performed on the sample with 15\,nm thick SRO layer and an uncovered STO substrate from the same manufacturer for comparison.

\section{TDBS with and BS without time resolution}
\label{sec:tdbs-vs-bs}
In this work we have studied the damping of LA phonons using two variants of Brillouin scattering experiments: In the time-resolved TDBS measurements we excite a metallic transducer layer with an ultrashort optical pump pulse with central wavelength $\lambda\!=\!800$\,nm derived from a Ti-Sapphire amplifier system running at 5\,kHz (Spectra Physics). After a certain time delay $\tau$ we probe the sample with a white light continuum pulse generated in a sapphire plate~\cite{Bradler2009} and record the reflectivity change $\Delta R(\tau)\!=\!\frac{R(\tau)-R(\tau\!=\!0)}{R(\tau\!=\!0)}$ where $R(\tau)$ is the time dependent optical sample reflectivity and $R(\tau\!=\!0)$ the reflectivity of the unpumped sample before excitation~\cite{Bojahr2013,Bojahr2015,Ruello2015}. The metallic transducers SRO respectively LSMO absorb the energy of the laser pulse and expand quasi-instantaneously. Subsequently, a longitudinal strain wave is launched into the STO substrate~\cite{Ruello2015}. This LA phonon propagates perpendicular to the sample surface, in our case along the $[100]$ direction of STO. The generated strain amplitudes have previously been calibrated for SRO and LSMO transducers under comparable conditions using ultrafast x ray diffraction: Strain amplitudes on the order of $\sim\!0.5$\% are typically generated~\cite{Bojahr2012a}.

\begin{figure}[!ht]
\centering
\includegraphics[width = \columnwidth]{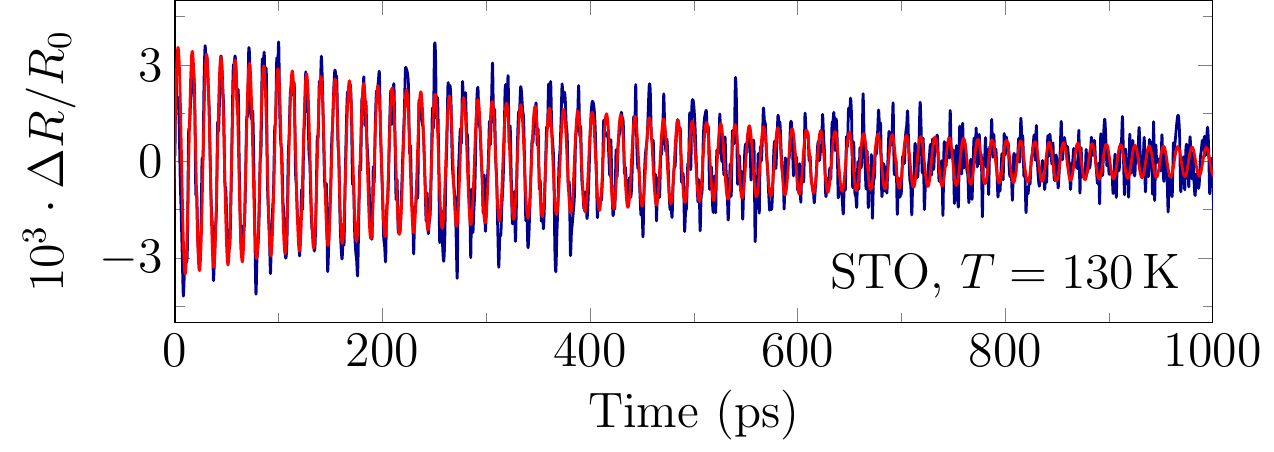}
\caption{Reflectivity change measured by TDBS at $T\!=\!130$\,K and 514\,nm using a single pulse laser excitation with fluence 14\,mJ/cm$^{2}$. The acoustic phonon is damped out within $\sim\!1$\,ns, which corresponds to a propagation length of $\sim\!8\,\mu$m.}\label{fig:TDBStransient}
\end{figure}

In Fig.~\ref{fig:TDBStransient} we show a typical TDBS measurement for $T\!=\!130\,\text{K}\!>\!T_{\text{a}}$, extracted at 514\,nm after optical excitation with $\lambda\!=\!800$\,nm and a fluence of 14\,mJ/cm$^{2}$. 
A damped cosine function has been fitted to the measured data as indicated by the red curve in Fig.~\ref{fig:TDBStransient}, which allows us to directly extract the damping rate $\Gamma$ of the acoustic phonon. In contrast to neutron and Brillouin scattering where the life time is inferred from the line width~\cite{Cowley1968}, in TDBS the damping rate is directly observed in the time domain. The phonon oscillation frequency of the excited acoustic phonon is seen as oscillating signal in the TDBS data. \footnote{The phonon frequency is given by $\nu(q) = {2v_{\text L}n(\lambda)\cos\theta}\!/\!_{\lambda}$ with sound velocity $v_{\text L}$, the wavelength dependent optical refractive index of STO $n(\lambda)$ taken from Ref.~\cite{Cardona1965}, and the angle $\theta$ between surface normal and angle of incidence of the probe light.} 
Monitoring the full white light continuum with a fiber-coupled Czerny-Turner spectrometer with CCD array (Avantes) allows us to simultaneously record a broad spectrum of the coherently generated phonons~\cite{Bojahr2012a,Bojahr2013,Maerten2015b} with wavevectors $45\,\mu\text{m}^{-1}\!<\!q\!<\!65\,\mu\text{m}^{-1}$ in the visible spectral region. 

\begin{figure}[!ht]
\centering
\includegraphics[width = \columnwidth]{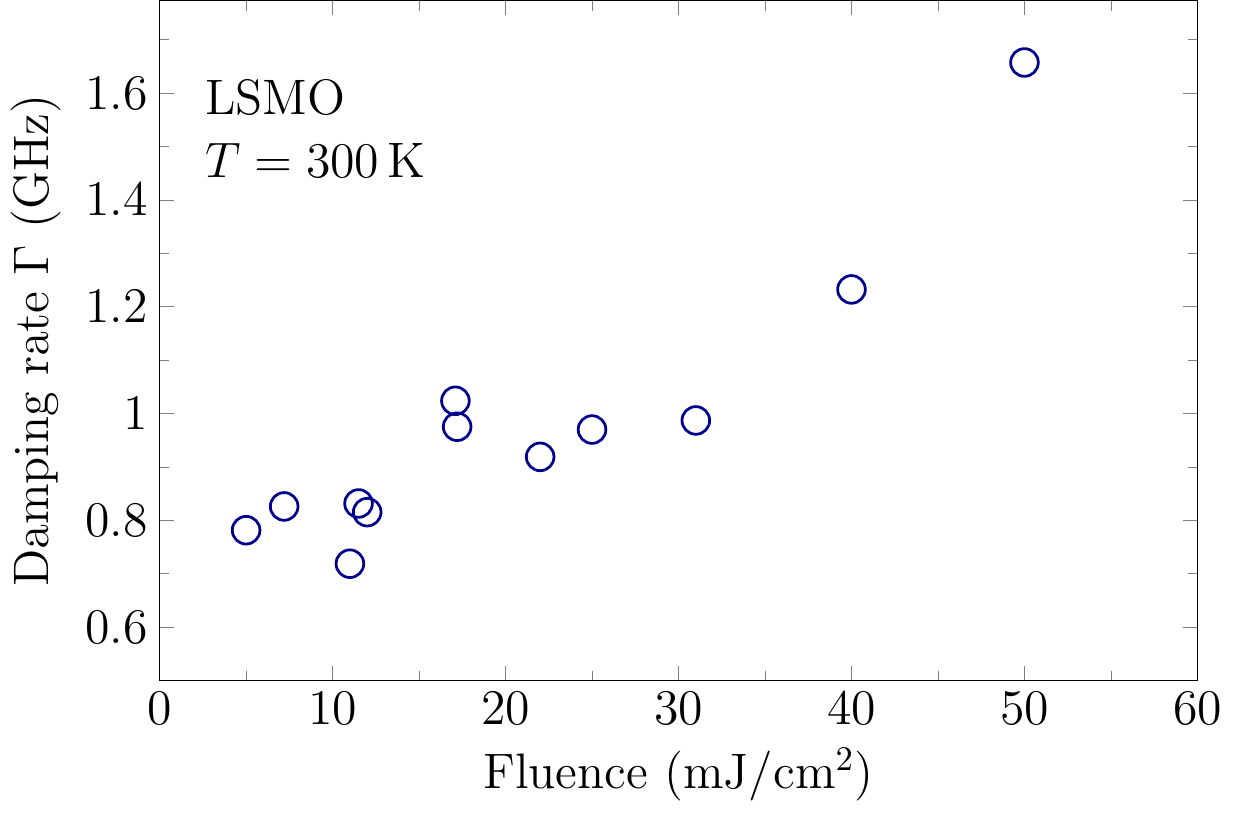}
\caption{Fluence-dependence of the damping rate $\Gamma$ for the multipulse excitation of the sample with LSMO transducer as determined at room temperature.}
\label{fig:f_dep_damping}
\end{figure}

Some of the authors have previously studied the non-linear LA phonon propagation in STO due to its strongly anharmonic lattice potential~\cite{Bojahr2012a,Bojahr2015}. Depending on the optical excitation amplitude, the anharmonicity of the lattice has to be taken into account. We show in Fig.~\ref{fig:f_dep_damping} the fluence dependence of $\Gamma$ at $T\!=\!300$\,K for the LSMO sample after multipulse excitation as a measure for the lattice anharmonicity. We have used a multipulse excitation where we distribute the energy of the excitation over several pulses and hence reduce the contribution of the anharmonic lattice contribution~\cite{Bojahr2015} and at the same time also excite narrow-bandwidth phonons with a well-defined $q$ vector~\cite{Bojahr2013}. For some of the experiments presented in this paper, we have used 8 pump pulses with a duration of $\sim\!7$\,ps and a temporal spacing of $\sim\!15$\,ps.
This excitation scheme generates a acoustic phonon with $q\approx 52\;\mu\text{m}^{-1}$ and we will compare in the following these results with the high fluence single pulse excitation experiments performed on the same samples and the ``static'' Brillouin scattering experiments. From the observed fluence dependence of $\Gamma$ after multipulse excitation we distinguish two different regimes: First, a low fluence range up to $\sim\!30$\,mJ/cm$^2$ where $\Gamma$ increases only slightly. The second regime for fluences higher than 30\,mJ/cm$^2$ suggests that the non-linear response of the lattice becomes important, which we associate with the steep increase of $\Gamma$ observed for the high fluences close to the damage threshold of the transducer. These high fluences are in the range where we have previously observed the superelastic effect for GHz phonons using a single pulse excitation of the LSMO sample~\cite{Maerten2015}.

\begin{figure}[!ht]
\centering
\includegraphics[width = \columnwidth]{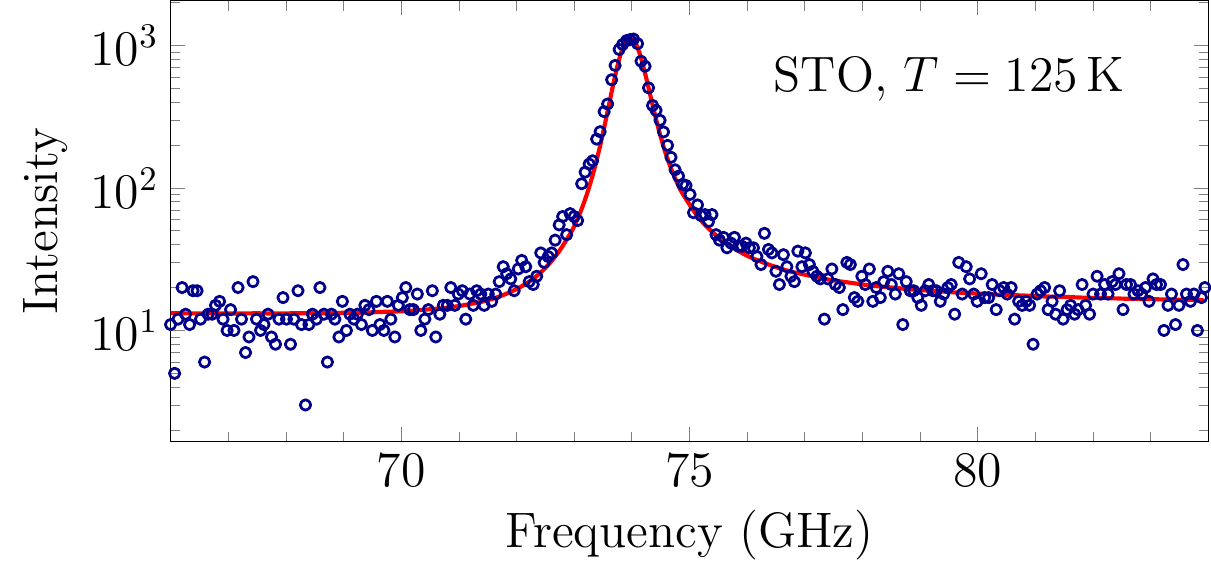}
\caption{Typical example of a frequency-resolved BS spectrum at $T\!=\!125$\,K. The red line is the fit to the experimental data using a Fano function convoluted with the instrument function. It yields the line width of $\beta\!=\!\Gamma/2\pi$.}
\label{fig:BS-spectrum}
\end{figure}

The ``static'' Brillouin scattering (BS) experiment on the other side measures the frequency shift of a continuous wave Nd:YAG laser with $\lambda\!=\!532$\,nm monitored in transmission geometry at normal incidence using a high resolution offset stabilized tandem (Sandercock) Fabry-P\'erot interferometer~\cite{Koreeda2011}. Consequently, phonons with $q\!=\!58\,\mu\text{m}^{-1}$ are probed. In Fig.~\ref{fig:BS-spectrum} a typical frequency resolved BS spectrum for $T\!>\!T_{\text{a}}$ is shown. 
The asymmetry of the line shape stems from the interference between the polarizations corresponding to the Brillouin and quasi-elastic scattering contributions~\cite{Koreeda2006}, although the detailed mechanism is still unclear. A Fano function~\cite{Fano1961} convoluted with the resolution function as indicated by the red line in Fig.~\ref{fig:BS-spectrum} accounts for the asymmetry and yields a good fit to the experimental data. A comparison of the damping rate deduced from the line width of the BS spectrum and the decaying envelope of the oscillation in the TDBS spectra is presented in Sec.~\ref{sec:results-discussion}.

\section{Structural sample characterization}
\label{sec:structural-characterization}
\begin{figure}
\centering\includegraphics[width = .4\textwidth]{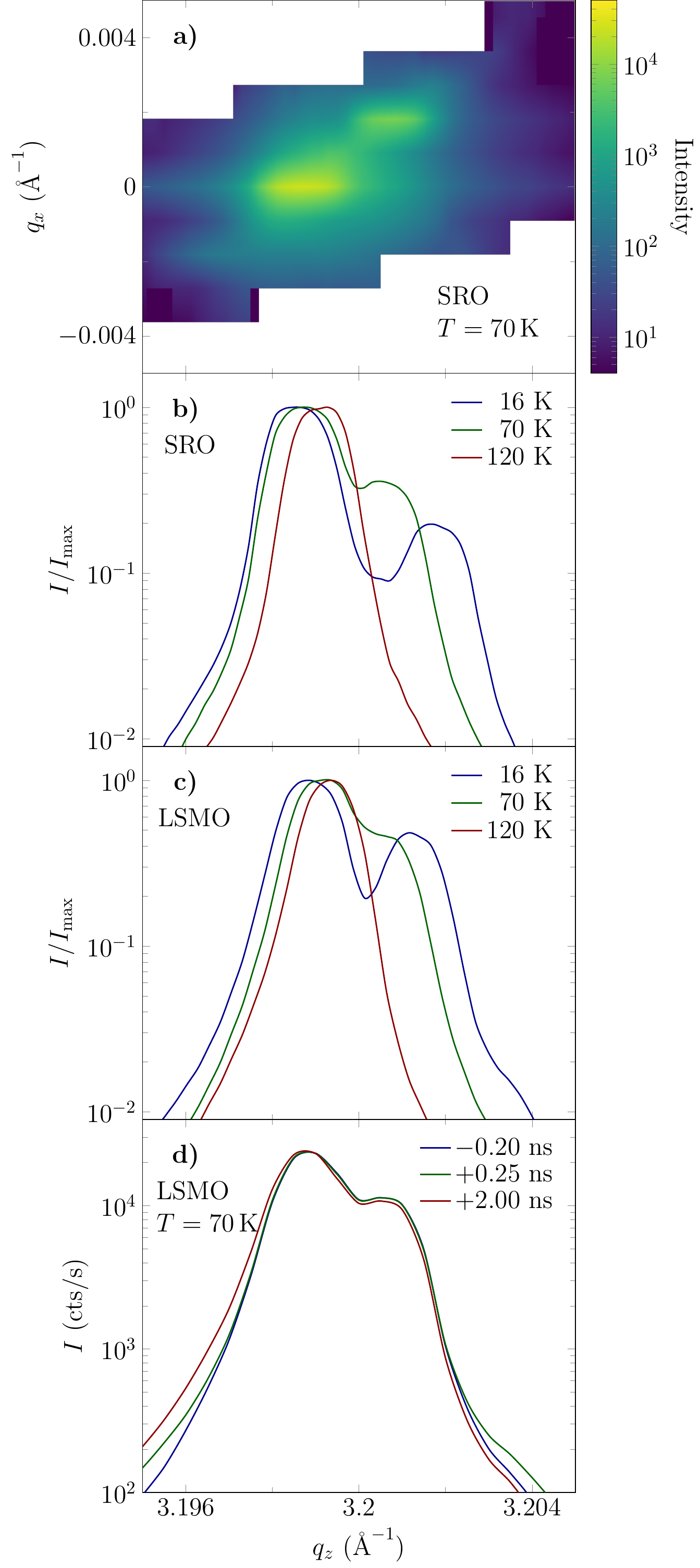}
\caption{a) RSM of the STO sample with SRO transducer at $T\!=\!70$\,K. The intense peak corresponds to the 002 Bragg reflection from domains with the elongated $c$ axis perpendicular to the surface whereas the smaller peak originates from the 200 and 020 reflections, that is, from $a$ or $b$ domains with the shorter axes perpendicular to the surface. Note that the secondary maxima are displaced in $q_z$ and $q_x$ of reciprocal space, thus confirming the domain structure proposed in Fig.~4 of Ref.~\cite{Maerten2015}. b) Projections of the RSM onto the $q_z$ axis around the 002 reflection of STO for the sample with an SRO transducer at different temperatures. c) Same for the STO sample with an LSMO transducer. d) Results from ultrafast RSM experiments at $T\!=\!70$\,K taking snapshots at 0.20\,ns before and 0.25 and 2\,ns after the optical excitation of the LSMO transducer. At 0.25\,ns shoulders on both sides of the Bragg peak originating from the coherent sound wave in STO are visible whereas at 2\,ns the soundwave is fully damped and only thermal heating of STO is seen as Bragg peak shift towards lower values of $q_z$.}
\label{fig:rsm-xray}
\end{figure}

In the following we confirm that the domain patterns of the STO substrates in the samples with SRO and LSMO transducers at temperatures $T\!<\!T_{\text{a}}$ are different.
Reciprocal space mapping (RSM) performed at the KMC3-XPP endstation of the synchrotron radiation source BESSY II, Berlin, Germany, using highly collimated hard x rays with photon energy 10\,keV probes the interfacial region between STO and the transducer layers. The penetration depth of 10\,keV photons is on the order of 2\,$\mu$m in STO for the 002 Bragg reflection. Hence, we probe preferably the interface between the metallic transducers and the STO substrates, similarly to the experiments discussed in Ref.~\cite{Loetzsch2010}. A typical RSM with $q_z$ and $q_x$ components is shown in Fig.~\ref{fig:rsm-xray}a) for the SRO sample at $T\!=\!70$\,K and in Figs.~\ref{fig:rsm-xray}b) and c) 
we show the corresponding normalized diffraction intensities $^I\!/\!_{I_{\text{max}}}$ for the SRO and LSMO samples, respectively, at three selected $T$ after the integration of the measured RSM intensities along $q_z$. We observe in the both samples below $T_{\text{a}}$ a splitting of the 002 Bragg reflection of STO into two components. We assign the more intense peak to domains with the tetragonal $c$ axis aligned perpendicular to the interface and the weaker peak to domains with their tetragonal axis oriented parallel to the interface. These results show that at low $T$ the splitting of the $a$ and $c$ domains is different for the SRO and the LSMO samples: For SRO we obtain $^c\!/\!_a\!=\!9\cdot10^{-4}$ whereas the LSMO sample exhibits a 20\% larger ratio of $^c\!/\!_a\!=\!7\cdot 10^{-4}$, which both are significantly larger than the tetragonality observed in bulk STO samples~\cite{Lytle1964}. In our opinion this explains why superelasticity for hypersound waves was only observed in samples with LSMO transducers since the suggested process in Ref.~\cite{Maerten2015} requires an overcompensation of the tetragonal distortion, and potentially, the required stress is above the damage threshold of the SRO transducer.
 
For completeness, in Fig.~\ref{fig:rsm-xray}d) we report ultrafast x ray diffraction data on the LSMO sample. The sample was at $T\!=\!70$\,K excited with a laser fluence of about 5\,mJ/cm$^2$ at the excitation wavelength $\lambda\!=\!1030$\,nm, that is, considerably less than the 45\,mJ/cm$^2$ required to observe the superelastic effect~\cite{Maerten2015}. Unfortunately, such high fluences are not available from the laser system at the KMC3-XPP endstation because the efficient synchronization of the laser pulses to the ring frequency favours high repetition rates of the laser pulses, which in turn implies low pulse energies. The green line in Fig.~\ref{fig:rsm-xray}d) shows that the Bragg peak develops shoulders at its high and low $q_z$ sides, indicating that a acoustic phonon is generated and propagates in STO 0.25\,ns after the excitation. Around 2\,ns, the sound wave is fully damped, which is consistent with our TDBS data shown in Fig.~\ref{fig:TDBStransient} where the sound wave is completely damped after $\sim\!1$\,ns. At later times, the Bragg peak shows the characteristic shift towards smaller $q_z$ values corresponding to a transient temperature increase because heat is diffusing from the laser-excited transducer into the substrate.

\section{Results and discussion}
\label{sec:results-discussion}

\begin{figure}[!ht]
\centering
\includegraphics[width = \columnwidth]{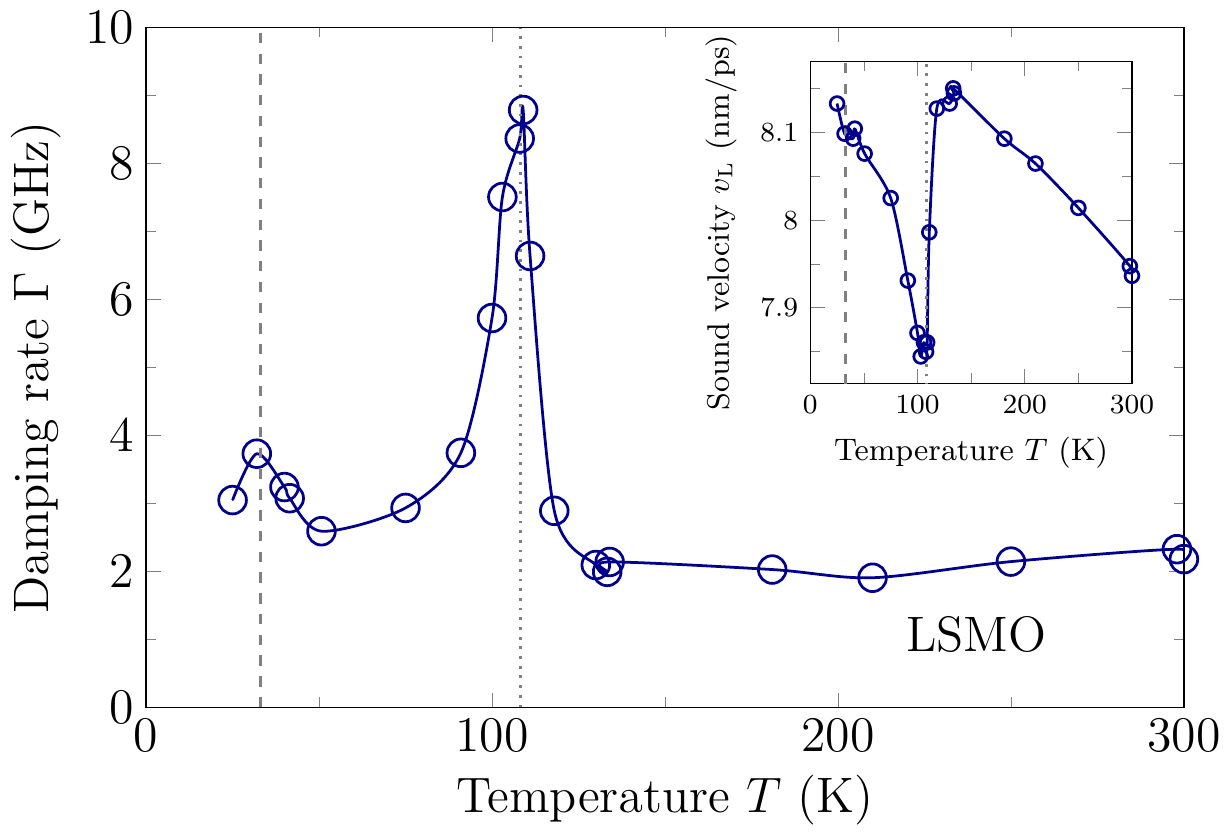}
\caption{Damping rate $\Gamma$ as function of $T$ for the single pulse excitation of the sample with LSMO transducer, excited with $\sim\!14$\,mJ/cm$^2$ and extracted at $\lambda\!=\!528$\,nm, which corresponds to $q\!=\!52\,\mu\text{m}^{-1}$. The inset displays the sound velocity, $v_L$, as function of $T$ for the same dataset, reproduced from Ref.~\cite{Maerten2015}. At $T_{\text{a}}\!=\!108.2$\,K, marked by the gray dotted vertical line, $\Gamma$ exhibits a maximum. The dashed vertical line at 33\,K marks an increase of $\Gamma$ that might be associated with an increased domain wall conductivity, see main text.}
\label{fig:damping_simple}
\end{figure}

Before we start the discussion of the results, we emphasize that TDBS probes the propagation of generated LA phonons~\cite{Bojahr2013} as the accompanying strain fields locally modify the refractive index of the surrounding material~\cite{Thomsen1986,Thomsen1986a,Bojahr2013}.
We extract from the TDBS spectra after a fast Fourier transformation of the measured TDBS intensities along the time axis for every probe wavelength, that is, every $q$ vector, the sound velocity $v_{\text L}$ of the excited LA phonon~\cite{Bojahr2012a}. The $T$ dependent sound velocity of STO as obtained from the experiment on the LSMO sample after single pulse excitation is shown in the inset of Fig.~\ref{fig:damping_simple}~\cite{Maerten2015}, consistent with the results in Ref.~\cite{Kaiser1966}. Simultaneously, as shown in Fig.~\ref{fig:TDBStransient}, the damping rate $\Gamma$ is obtained as the ``decay constant'' of the cosine fit function. The results of the evolution of $\Gamma$ as function of $T$ is displayed in the main panel of Fig.~\ref{fig:damping_simple}.  
At high $T$ in the cubic phase of STO, we observe a $T$ independent value of $\Gamma\!\approx\!2$\,GHz. As the temperature approaches $T_{\text{a}}$, $\Gamma$ steeply increases and reaches its maximum of $\Gamma\!\approx\!9$\,GHz at $T_{\text{a}}\!\approx\!108$\,K. As the temperature is further lowered, the damping rate decreases and settles at a slightly higher value of 2.5\,GHz at $\sim\!65$\,K. At $T\!\approx\!33$\,K, an additional increase of the damping rate is observed. At this $T$ the mechanical properties of STO also exhibit discontinuities~\cite{Bell1963,Binder2001,Carpenter2007}. However, recent works imply that the origin of the sound velocity reduction and hence the increase of the damping rate around $\sim\!33$\,K is caused by the crossing or at least strong interaction of the ferroelectric and the AFD soft phonon modes~\cite{Tagantsev2001}. It seems that polar domain walls are realized in STO below $\sim\!50$\,K~\cite{Scott2012a,Salje2013,Salje2016} and the associated mobility increase~\cite{Zykova-Timan2014} might be observed as increase of $\Gamma$ in our measurements.   

\begin{figure}[!ht]
\centering
\includegraphics[width = \columnwidth]{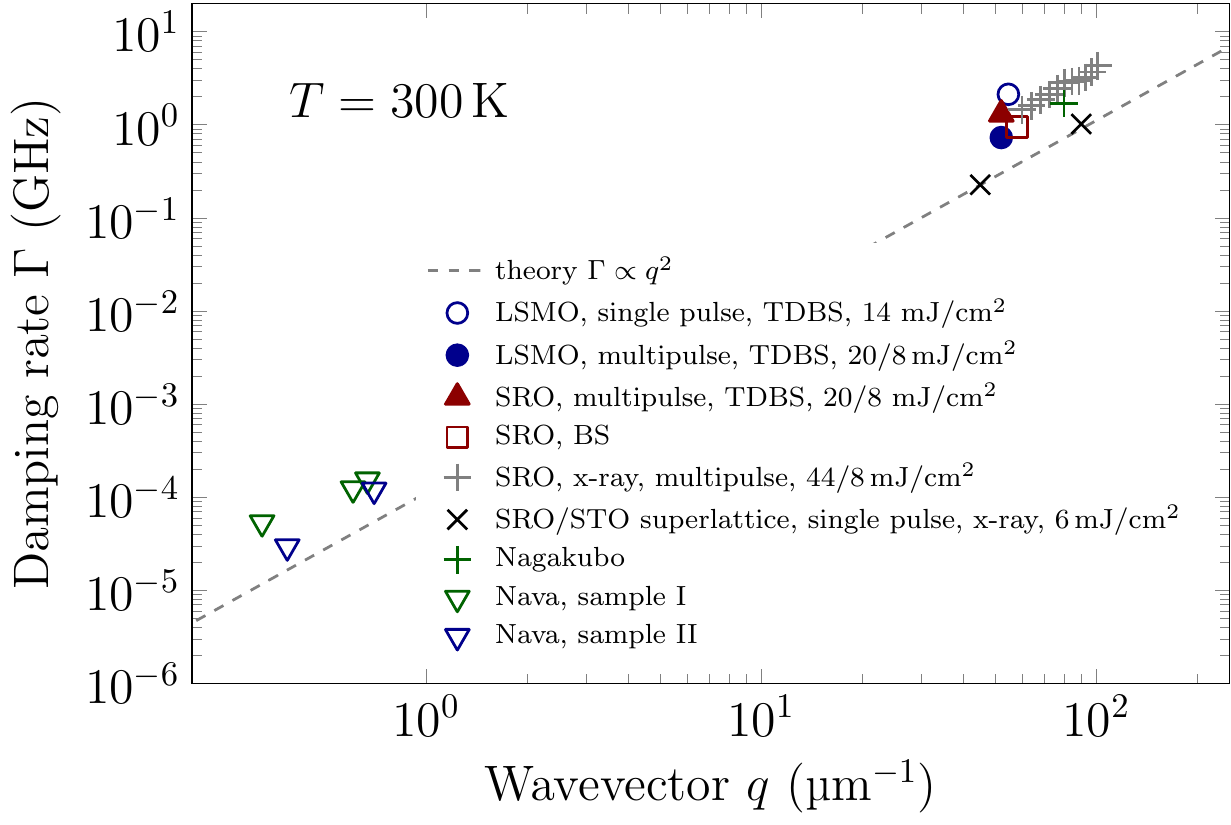}
\caption{Comparison of the observed damping rates at room temperature for various samples and with different measurement techniques with theoretical and literature values. X-ray data according to Herzog \textit{et~ al.} using pulse train excitation~\cite{Herzog2012a}, and Shayduk \textit{et~ al.}, exciting a superlattice of 5 double layers of SRO/STO on a STO substrate~\cite{Shayduk2013}. Literature data from Nava \textit{et~ al.}~\cite{Nava1969} and Nagakubo \textit{et~ al.}~\cite{Nagakubo2012} are included. The theoretically expected $q^2$ dependence shown as dashed line was calculated according to Akhiezer's model and literature values.}
\label{fig:masterplot}
\end{figure}

\begin{figure}[!ht]
\centering
\includegraphics[width = \columnwidth]{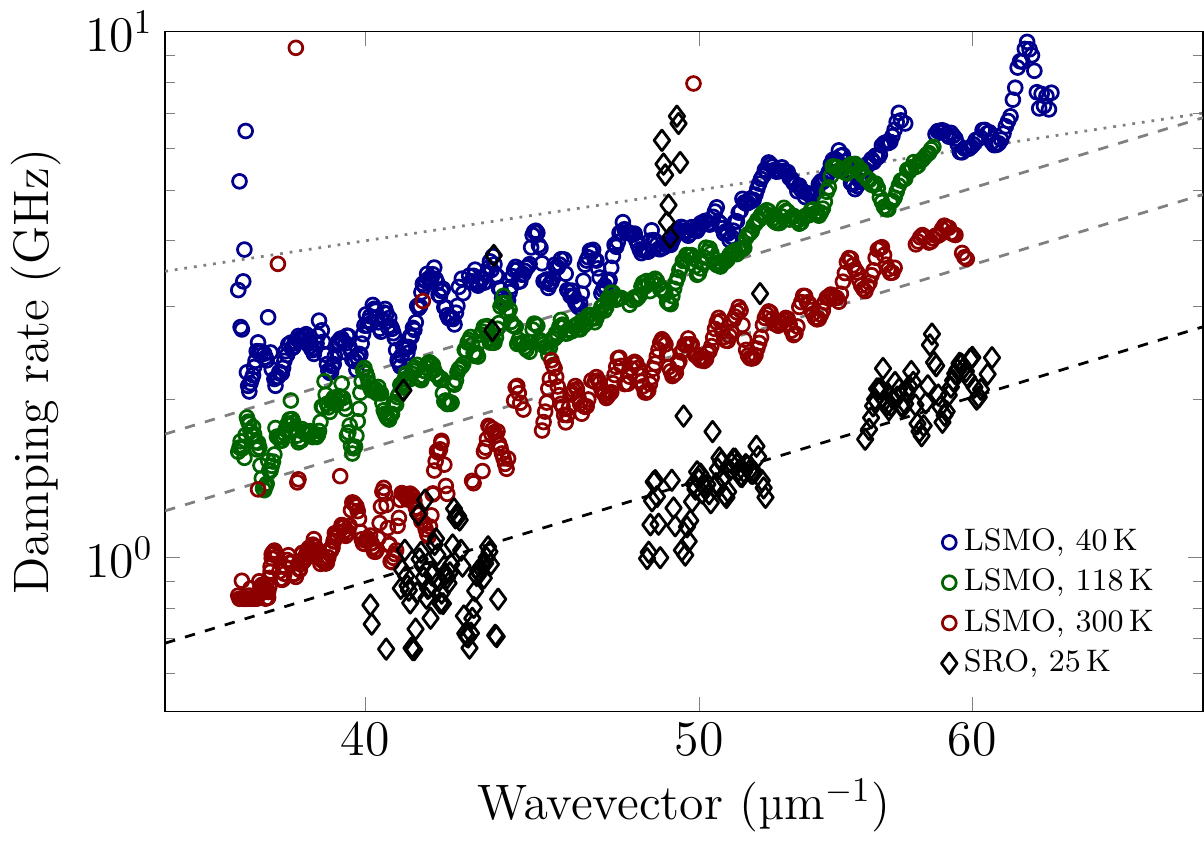}
\caption{$q$-dependence of the damping rate $\Gamma$ as extracted at different temperatures for the LSMO sample. The dotted line indicates a $q$- and the dashed lines a $q^2$-dependence of the damping rate according to Akhiezer's model of the phonon damping. We show for comparison the result for the SRO sample at $T = 25$\,K in black.}
\label{fig:q_dependence}
\end{figure}

An important aspect of the phonon damping rate is its $q$ dependence. In Fig.~\ref{fig:masterplot} we show our results and compare these with results from other measurements at different $q$ vectors, all measured at room temperature~\cite{Nava1969,Herzog2012a,Nagakubo2012,Shayduk2013}. The dashed line indicates a line with slope 2 as calculated from Akhiezer's model~\cite{Akhiezer1939}, who proposed a $q^2$ dependence of $\Gamma$. Our experimental results confirm this model for $T\!=\!300$\,K. 
As we decrease the temperature, we observe in the visible spectral region the same $q^2$ behavior as shown in Fig.~\ref{fig:q_dependence}.

\begin{figure*}
\centering
\includegraphics[width = \textwidth]{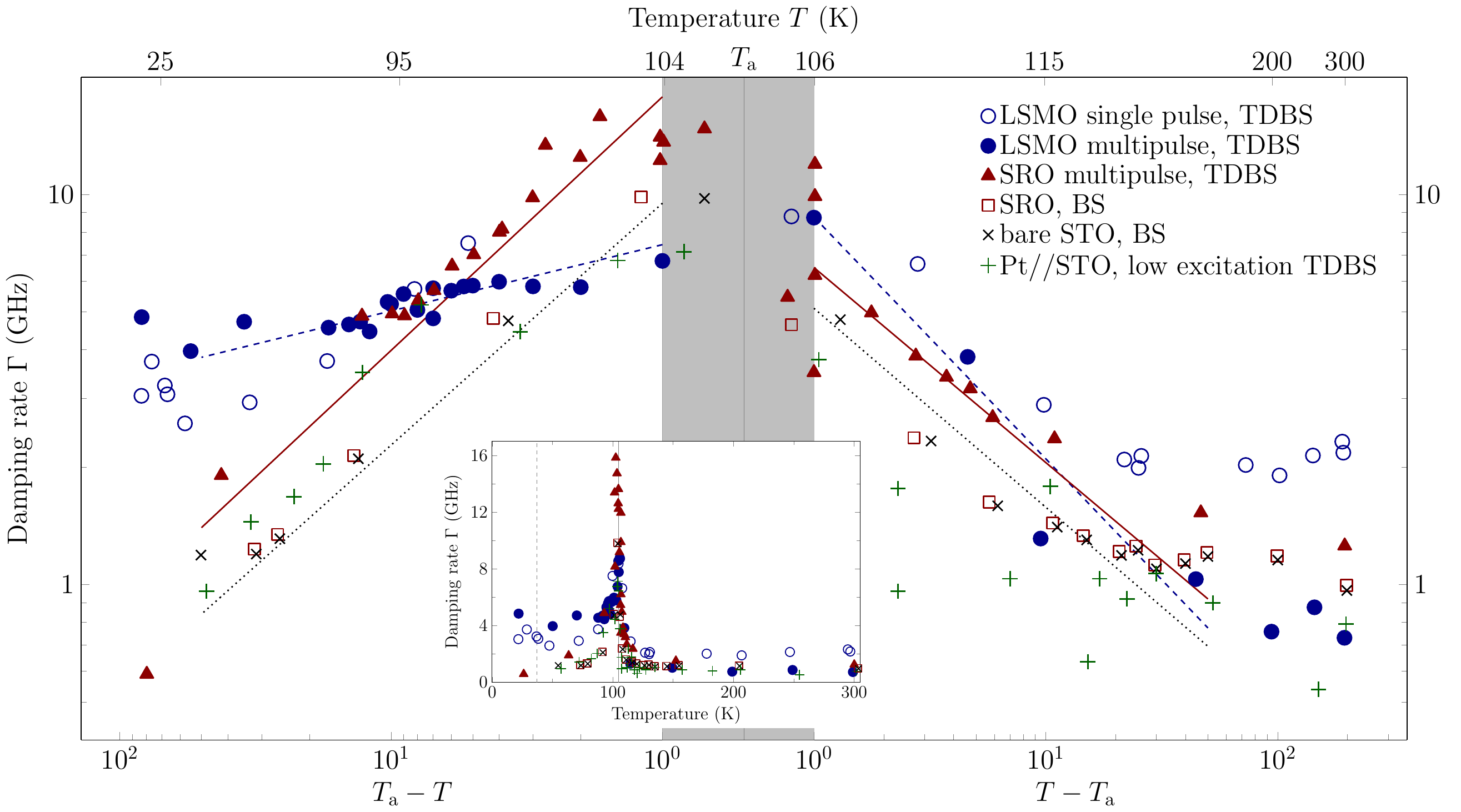}
\caption{Damping rate $\Gamma$ as function of the reduced temperature $|T-T_{\text{a}}|$ on a double logarithmic scale. We show the results for the LSMO sample, the SRO sample, and bare STO as obtained by TDBS after single pulse or multipulse excitation, and BS. For comparison, we show the rescaled results from Ref.~\cite{Nagakubo2012} after low power excitation of a Pt transducer. The inset contains the same data points on a linear scale. The dashed lines are exemplary fits to the damping rates in order to extract the critical exponents $\eta'$ and $\eta$ of the phase transition below and above $T_{\text{a}}$, respectively. For a better visual comparability, all measurements have been shifted such that the maximum damping rate occurs at $T_{\text{a}}\!=\!105$\,K, the transition temperature of bulk STO. The experimentally determined values of $T_{\text{a}}$ from our measurements and the results of the fits are given in Tab.~\ref{tab:ta-damping}.}
\label{fig:damping_all}
\end{figure*}

Now we proceed to the discussion of the $T$ dependence of $\Gamma$. In Fig.~\ref{fig:damping_all} we display the extracted damping rates for all samples, measurement, and excitation schemes as function of $|T-T_{\text{a}}|$ on a double logarithmic plot and in the inset on a linear scale~\footnote{In the linear inset we shifted the maxima of $\Gamma(T)$ to $T_{\text{a}}\!=\!105$\,K. We have obtained the ``exact'' transition temperatures by the fit using the function $A\cdot|T - T_{\text{a}}|^{-\eta}$ which yields the slope $\eta$ of the linear fit together with a fit constant $A$, which represents the offset of the abscissa. The values of $\eta$ ($\eta'$) and $T_{\text{a}}$ are given in Tab.~\ref{tab:ta-damping}.}. We compare here single and multipulse excitations for the TDBS measurements with the continuous wave excitation during the BS experiments and observe distinctive differences of the absolute value of $\Gamma$ and also different $T$ dependencies. Already at room temperature we note the following differences between the measured samples: The phonon damping is in our samples at 300\,K on the order of 1-2\,GHz. The single pulse, high fluence TDBS measurements result in higher damping rates of $\sim\!2$\,GHz, the multipulse excitation measurements yield together with the low excitation TDBS measurement from Ref.~\cite{Nagakubo2012} and the BS experiments systematically lower values of $\sim\!1$\,GHz.
Significant differences between samples and measurements can be identified. In our experiments the highest value $\Gamma\!\approx\!16$\,GHz occurs in the SRO sample after multipulse excitation, the other experiments and samples typically show damping rates on the order of 8-10\,GHz. In the low $T$ tetragonal phase, the LSMO measurements indicate for single and multipulse excitation significantly higher damping rates of $\sim\!4$\,Ghz compared to the SRO and bulk STO results of $\sim\!1$\,GHz.

We propose that the $T$ dependence of $\Gamma$ exhibits also for GHz LA phonons a critical behavior. Hence, we expect a linear behavior of $\Gamma$ as function of the reduced temperature $|T-T_{\text{a}}|$ in a double logarithmic representation. From our measurements we cannot determine $T_{\text{a}}$ with an accuracy better than $\pm\!1$\,K. We visualize this uncertainty range by the gray shaded area in Fig.~\ref{fig:damping_all}. The uncertainty originates dominantly from the fact that $\Gamma$ rises very sharply close to $T_{\text{a}}$ as can clearly be seen in the linearly scaled inset of Fig.~\ref{fig:damping_all}. Thus, already small temperature variations have a large influence on the extracted values of $\Gamma$ in the vicinity of $T_{\text{a}}$. In addition, the TDBS and BS measurements have been performed with different setups, which gives systematic uncertainties in reading the temperature. 
We note that furthermore $T_{\text{a}}$ can be increased by pressure~\cite{Fossheim1972,Guennou2010} or reduced due to a possible charge carrier accumulation, which can for example be the result of an oxygen non-stoichiometry~\cite{Bauerle1978a}. We exemplary show in Fig.~\ref{fig:damping_all} linear fits for LSMO TDBS data with multipulse excitation (dashed blue), the SRO TDBS measurement with multipulse excitation (solid red), and for the BS results of bare STO (dotted black). The resulting critical exponents $\eta$ for $T\!>\!T_{\text{a}}$ ($\eta'$ for $T\!<\!T_{\text{a}}$) as well as the values of $T_{\text a}$ are summarized in Tab.~\ref{tab:ta-damping}.
\begin{table}
\caption{\label{tab:ta-damping} Experimentally determined values of $T_{\text{a}}$ and results of the linear fits to the damping rate in Fig.~\ref{fig:damping_all} for the different samples and measurement schemes. $\eta'$ are the exponents for $T\!<\!T_{\text{a}}$ and $\eta$ for $T\!>\!T_{\text{a}}$.}
\begin{ruledtabular}
\begin{tabular}{llll}
sample, excitation scheme & $T_{\text{a}}$~(K) & $\eta'$ & $\eta$\\
\hline
LSMO single pulse, TDBS & 108.2 & $0.44\pm0.06$ & $0.56\pm0.04$ \\
LSMO multipulse, TDBS & 106.0 & $0.17\pm0.03$ & $0.62\pm0.22$ \\
SRO multipulse, TDBS & 104.1 & $0.65\pm0.19$ & $0.50\pm0.03$ \\
SRO, BS & 100.5 & $0.61\pm0.02$ & $0.47\pm0.02$ \\
STO, BS & 100.0 & $0.62\pm0.04$ & $0.51\pm0.05$ \\
Pt//STO, TDBS & 97.8 & $0.46\pm0.14$ & $0.38\pm0.10$ \\
\end{tabular}
\end{ruledtabular}
\end{table}
In the cubic phase above $T_{\text{a}}$, the fits describe the experimental results in the temperature interval $T_{\text{a}}\!\leq\!T_{\text{a}}+40$\,K in agreement with the data and results presented in Ref.~\cite{VonWaldkirch1972}. In the tetragonal phase below $T_{\text{a}}$, we observe a linear dependence of $\Gamma$ as function of the reduced temperature down to $T_{\text{a}}\!\geq\!T_{\text{a}}-70$\,K.

We discuss now the values of $\eta$ and $\eta'$ that we obtain for the different samples and excitation schemes. In the literature, one finds a large variation of the reported values of the critical exponents in STO and measured by different experimental techniques. Our results yield the critical dynamical exponent for the LA phonon damping rate, which is of course inversely proportional to the phonon attenuation.
Our results for $\eta$, that is, for $T\!>\!T_{\text{a}}$, agree with results predicted by Rehwald~\cite{Rehwald1971} at MHz frequencies, closely resembling the mean-field value of $\eta$. However, they differ significantly from published results for MHz phonons~\cite{Berre1969,Fossheim1972} where $\eta\approx1.25$ has been determined. 

At $T\!<\!T_{\text{a}}$, we observe different values of the critical exponent $\eta'$ than we have observed above $T_{\text{a}}$. In case of the LSMO sample, we extract lower values of $\eta'$ for the single pulse excitation and a significantly reduced value after the excitation with multiple pulses. In contrast to this behavior, we yield for SRO and bare STO samples higher values for $\eta'$ than for $\eta$. We attribute these differences to the different domain structures of the samples below $T_{\text a}$: The critical coefficients of the phonon damping rate reflect the fact that the motion of the AFD twin domain walls under compressive (tensile) strain are damped out less (more) efficiently considering the comparable multipulse excitation. This is even more surprising as the fluence per pulse is $^{20}\!/\!_{8}\!=\!4.5\,\text{mJ/cm}^2$, a value that is far away from the threshold where we have observed the superelastic effect in the single pulse excitation. 

\section{Conclusions}
We have measured the damping rate of GHz acoustic phonons in SrTiO$_3$. Our time-resolved and frequency-resolved Brillouin scattering experiments indicate particular differences in the damping rate for bulk SrTiO$_3$ as compared to SrTiO$_3$ samples with epitaxially grown thin metallic transducer layers that exert in-plane compressive or tensile strain, respectively. Our reciprocal space maps indicate that the compressively strained samples with LSMO transducers exhibit a different domain pattern than the samples under tensile strain. This manifests itself in different damping rate of the GHz longitudinal acoustic phonons below the antiferrodistortive phase transition temperature of SrTiO$_3$. We have furthermore shown that the the phonon damping rate exhibits a critical behavior with different critical exponents below and above the phase transition.

\acknowledgements
We thank I. Vrejoiu for providing the samples and the DFG for support via grant No. BA~2281/8-1. AB thanks the Leibniz Graduate school DinL for financial support.

\section{References}


\begin{thebibliography}{60}
\expandafter\ifx\csname natexlab\endcsname\relax\def\natexlab#1{#1}\fi
\expandafter\ifx\csname bibnamefont\endcsname\relax
  \def\bibnamefont#1{#1}\fi
\expandafter\ifx\csname bibfnamefont\endcsname\relax
  \def\bibfnamefont#1{#1}\fi
\expandafter\ifx\csname citenamefont\endcsname\relax
  \def\citenamefont#1{#1}\fi
\expandafter\ifx\csname url\endcsname\relax
  \def\url#1{\texttt{#1}}\fi
\expandafter\ifx\csname urlprefix\endcsname\relax\def\urlprefix{URL }\fi
\providecommand{\bibinfo}[2]{#2}
\providecommand{\eprint}[2][]{\url{#2}}

\bibitem[{\citenamefont{M{\"{u}}ller and Burkhard}(1979)}]{Muller1979}
\bibinfo{author}{\bibfnamefont{K.~A.} \bibnamefont{M{\"{u}}ller}}
  \bibnamefont{and} \bibinfo{author}{\bibfnamefont{H.}~\bibnamefont{Burkhard}},
  \bibinfo{journal}{Phys. Rev. B} \textbf{\bibinfo{volume}{19}},
  \bibinfo{pages}{3593} (\bibinfo{year}{1979}).

\bibitem[{\citenamefont{Lytle}(1964)}]{Lytle1964}
\bibinfo{author}{\bibfnamefont{F.~W.} \bibnamefont{Lytle}},
  \bibinfo{journal}{J. Appl. Phys.} \textbf{\bibinfo{volume}{35}},
  \bibinfo{pages}{2212} (\bibinfo{year}{1964}).

\bibitem[{\citenamefont{Shirane and Yamada}(1969)}]{Shirane1969}
\bibinfo{author}{\bibfnamefont{G.}~\bibnamefont{Shirane}} \bibnamefont{and}
  \bibinfo{author}{\bibfnamefont{Y.}~\bibnamefont{Yamada}},
  \bibinfo{journal}{Phys. Rev.} \textbf{\bibinfo{volume}{177}},
  \bibinfo{pages}{858} (\bibinfo{year}{1969})

\bibitem[{\citenamefont{Sidoruk et~al.}(2010)\citenamefont{Sidoruk, Leist,
  Gibhardt, Meven, Hradil, and Eckold}}]{Sidoruk2010}
\bibinfo{author}{\bibfnamefont{J.}~\bibnamefont{Sidoruk}},
  \bibinfo{author}{\bibfnamefont{J.}~\bibnamefont{Leist}},
  \bibinfo{author}{\bibfnamefont{H.}~\bibnamefont{Gibhardt}},
  \bibinfo{author}{\bibfnamefont{M.}~\bibnamefont{Meven}},
  \bibinfo{author}{\bibfnamefont{K.}~\bibnamefont{Hradil}}, \bibnamefont{and}
  \bibinfo{author}{\bibfnamefont{G.}~\bibnamefont{Eckold}},
  \bibinfo{journal}{J. Phys. Condens. Matter} \textbf{\bibinfo{volume}{22}},
  \bibinfo{pages}{235903} (\bibinfo{year}{2010})

\bibitem[{\citenamefont{Cowley}(1996)}]{Cowley1996a}
\bibinfo{author}{\bibfnamefont{R.}~\bibnamefont{Cowley}},
  \bibinfo{journal}{Philos. Trans. Math. Phys. Eng. Sci.}
  \textbf{\bibinfo{volume}{354}}, \bibinfo{pages}{2799} (\bibinfo{year}{1996}),

\bibitem[{\citenamefont{Carpenter}(2007)}]{Carpenter2007}
\bibinfo{author}{\bibfnamefont{M.~A.} \bibnamefont{Carpenter}},
  \bibinfo{journal}{Am. Mineral.} \textbf{\bibinfo{volume}{92}},
  \bibinfo{pages}{309} (\bibinfo{year}{2007}).

\bibitem[{\citenamefont{Perks et~al.}(2014)\citenamefont{Perks, Zhang,
  Harrison, and Carpenter}}]{Perks2014}
\bibinfo{author}{\bibfnamefont{N.~J.} \bibnamefont{Perks}},
  \bibinfo{author}{\bibfnamefont{Z.}~\bibnamefont{Zhang}},
  \bibinfo{author}{\bibfnamefont{R.~J.} \bibnamefont{Harrison}},
  \bibnamefont{and} \bibinfo{author}{\bibfnamefont{M.~A.}
  \bibnamefont{Carpenter}}, \bibinfo{journal}{J. Phys. Condens. Matter}
  \textbf{\bibinfo{volume}{26}}, \bibinfo{pages}{505402}

\bibitem[{\citenamefont{Buckley et~al.}(1999)\citenamefont{Buckley, Rivera, and
  Salje}}]{Buckley1999}
\bibinfo{author}{\bibfnamefont{A.}~\bibnamefont{Buckley}},
  \bibinfo{author}{\bibfnamefont{J.~P.} \bibnamefont{Rivera}},
  \bibnamefont{and} \bibinfo{author}{\bibfnamefont{E.~K.~H.}
  \bibnamefont{Salje}}, \bibinfo{journal}{J. Appl. Phys.}
  \textbf{\bibinfo{volume}{86}}, \bibinfo{pages}{1653} (\bibinfo{year}{1999}),

\bibitem[{\citenamefont{Salje}(2012)}]{Salje2012}
\bibinfo{author}{\bibfnamefont{E.~K.} \bibnamefont{Salje}},
  \bibinfo{journal}{Annu. Rev. Mater. Res.} \textbf{\bibinfo{volume}{42}},
  \bibinfo{pages}{265} (\bibinfo{year}{2012})

\bibitem[{\citenamefont{Honig et~al.}(2013)\citenamefont{Honig, Sulpizio,
  Drori, Joshua, Zeldov, and Ilani}}]{Honig2013}
\bibinfo{author}{\bibfnamefont{M.}~\bibnamefont{Honig}},
  \bibinfo{author}{\bibfnamefont{J.~A.} \bibnamefont{Sulpizio}},
  \bibinfo{author}{\bibfnamefont{J.}~\bibnamefont{Drori}},
  \bibinfo{author}{\bibfnamefont{A.}~\bibnamefont{Joshua}},
  \bibinfo{author}{\bibfnamefont{E.}~\bibnamefont{Zeldov}}, \bibnamefont{and}
  \bibinfo{author}{\bibfnamefont{S.}~\bibnamefont{Ilani}},
  \bibinfo{journal}{Nat. Mater.} \textbf{\bibinfo{volume}{12}},
  \bibinfo{pages}{1112} (\bibinfo{year}{2013})

\bibitem[{\citenamefont{Kalisky et~al.}(2013)\citenamefont{Kalisky, Spanton,
  Noad, Kirtley, Nowack, Bell, Sato, Hosoda, Xie, Hikita et~al.}}]{Kalisky2013}
\bibinfo{author}{\bibfnamefont{B.}~\bibnamefont{Kalisky}},
  \bibinfo{author}{\bibfnamefont{E.~M.} \bibnamefont{Spanton}},
  \bibinfo{author}{\bibfnamefont{H.}~\bibnamefont{Noad}},
  \bibinfo{author}{\bibfnamefont{J.~R.} \bibnamefont{Kirtley}},
  \bibinfo{author}{\bibfnamefont{K.~C.} \bibnamefont{Nowack}},
  \bibinfo{author}{\bibfnamefont{C.}~\bibnamefont{Bell}},
  \bibinfo{author}{\bibfnamefont{H.~K.} \bibnamefont{Sato}},
  \bibinfo{author}{\bibfnamefont{M.}~\bibnamefont{Hosoda}},
  \bibinfo{author}{\bibfnamefont{Y.}~\bibnamefont{Xie}},
  \bibinfo{author}{\bibfnamefont{Y.}~\bibnamefont{Hikita}},
  \bibnamefont{et~al.}, \bibinfo{journal}{Nat. Mater.}
  \textbf{\bibinfo{volume}{12}}, \bibinfo{pages}{1091} (\bibinfo{year}{2013}),

\bibitem[{\citenamefont{Bell and Rupprecht}(1963)}]{Bell1963}
\bibinfo{author}{\bibfnamefont{R.}~\bibnamefont{Bell}} \bibnamefont{and}
  \bibinfo{author}{\bibfnamefont{G.}~\bibnamefont{Rupprecht}},
  \bibinfo{journal}{Phys. Rev.} \textbf{\bibinfo{volume}{129}},
  \bibinfo{pages}{90} (\bibinfo{year}{1963}),

\bibitem[{\citenamefont{Kaiser and Zurek}(1966)}]{Kaiser1966}
\bibinfo{author}{\bibfnamefont{W.}~\bibnamefont{Kaiser}} \bibnamefont{and}
  \bibinfo{author}{\bibfnamefont{R.}~\bibnamefont{Zurek}},
  \bibinfo{journal}{Phys. Lett.} \textbf{\bibinfo{volume}{23}},
  \bibinfo{pages}{668} (\bibinfo{year}{1966}),

\bibitem[{\citenamefont{Nava et~al.}(1969)\citenamefont{Nava, Callarotti, Ceva,
  and Martinet}}]{Nava1969}
\bibinfo{author}{\bibfnamefont{R.}~\bibnamefont{Nava}},
  \bibinfo{author}{\bibfnamefont{R.}~\bibnamefont{Callarotti}},
  \bibinfo{author}{\bibfnamefont{H.}~\bibnamefont{Ceva}}, \bibnamefont{and}
  \bibinfo{author}{\bibfnamefont{A.}~\bibnamefont{Martinet}},
  \bibinfo{journal}{Phys. Rev. B} \textbf{\bibinfo{volume}{188}},
  \bibinfo{pages}{1456} (\bibinfo{year}{1969}).

\bibitem[{\citenamefont{Berre et~al.}(1969)\citenamefont{Berre, Fossheim, and
  M{\"{u}}ller}}]{Berre1969}
\bibinfo{author}{\bibfnamefont{B.}~\bibnamefont{Berre}},
  \bibinfo{author}{\bibfnamefont{K.}~\bibnamefont{Fossheim}}, \bibnamefont{and}
  \bibinfo{author}{\bibfnamefont{K.~A.} \bibnamefont{M{\"{u}}ller}},
  \bibinfo{journal}{Phys. Rev. Lett.} \textbf{\bibinfo{volume}{23}},
  \bibinfo{pages}{589} (\bibinfo{year}{1969})

\bibitem[{\citenamefont{Tagantsev et~al.}(2001)\citenamefont{Tagantsev,
  Courtens, and Arzel}}]{Tagantsev2001}
\bibinfo{author}{\bibfnamefont{A.}~\bibnamefont{Tagantsev}},
  \bibinfo{author}{\bibfnamefont{E.}~\bibnamefont{Courtens}}, \bibnamefont{and}
  \bibinfo{author}{\bibfnamefont{L.}~\bibnamefont{Arzel}},
  \bibinfo{journal}{Phys. Rev. B} \textbf{\bibinfo{volume}{64}},
  \bibinfo{pages}{224107} (\bibinfo{year}{2001})

\bibitem[{\citenamefont{Kityk et~al.}(2000)\citenamefont{Kityk, Schranz, and
  Sondergeld}}]{Kityk2000}
\bibinfo{author}{\bibfnamefont{A.}~\bibnamefont{Kityk}},
  \bibinfo{author}{\bibfnamefont{W.}~\bibnamefont{Schranz}}, \bibnamefont{and}
  \bibinfo{author}{\bibfnamefont{P.}~\bibnamefont{Sondergeld}},
  \bibinfo{journal}{Phys. Rev. B} \textbf{\bibinfo{volume}{61}},
  \bibinfo{pages}{946} (\bibinfo{year}{2000}),

\bibitem[{\citenamefont{Schranz}(2011)}]{Schranz2011}
\bibinfo{author}{\bibfnamefont{W.}~\bibnamefont{Schranz}},
  \bibinfo{journal}{Phys. Rev. B} \textbf{\bibinfo{volume}{83}},
  \bibinfo{pages}{094120} (\bibinfo{year}{2011})

\bibitem[{\citenamefont{Maerten
  et~al.}(2015{\natexlab{a}})\citenamefont{Maerten, Bojahr, Gohlke,
  R{\"{o}}ssle, and Bargheer}}]{Maerten2015}
\bibinfo{author}{\bibfnamefont{L.}~\bibnamefont{Maerten}},
  \bibinfo{author}{\bibfnamefont{A.}~\bibnamefont{Bojahr}},
  \bibinfo{author}{\bibfnamefont{M.}~\bibnamefont{Gohlke}},
  \bibinfo{author}{\bibfnamefont{M.}~\bibnamefont{R{\"{o}}ssle}},
  \bibnamefont{and} \bibinfo{author}{\bibfnamefont{M.}~\bibnamefont{Bargheer}},
  \bibinfo{journal}{Phys. Rev. Lett.} \textbf{\bibinfo{volume}{114}},
  \bibinfo{pages}{047401} (\bibinfo{year}{2015}{\natexlab{a}}).

\bibitem[{\citenamefont{Salje et~al.}(2017)\citenamefont{Salje, Wang, Ding, and
  Scott}}]{Salje2017}
\bibinfo{author}{\bibfnamefont{E.~K.~H.} \bibnamefont{Salje}},
  \bibinfo{author}{\bibfnamefont{X.}~\bibnamefont{Wang}},
  \bibinfo{author}{\bibfnamefont{X.}~\bibnamefont{Ding}}, \bibnamefont{and}
  \bibinfo{author}{\bibfnamefont{J.~F.} \bibnamefont{Scott}},
  \bibinfo{journal}{Adv. Funct. Mater.} \textbf{\bibinfo{volume}{27}},
  \bibinfo{pages}{1700367} (\bibinfo{year}{2017})

\bibitem[{\citenamefont{R{\"{u}}tt et~al.}(1997)\citenamefont{R{\"{u}}tt,
  Diederichs, Schneider, and Shirane}}]{Rutt1997}
\bibinfo{author}{\bibfnamefont{U.}~\bibnamefont{R{\"{u}}tt}},
  \bibinfo{author}{\bibfnamefont{A.}~\bibnamefont{Diederichs}},
  \bibinfo{author}{\bibfnamefont{J.~R.} \bibnamefont{Schneider}},
  \bibnamefont{and} \bibinfo{author}{\bibfnamefont{G.}~\bibnamefont{Shirane}},
  \bibinfo{journal}{Eur. Lett.} \textbf{\bibinfo{volume}{39}},
  \bibinfo{pages}{395} (\bibinfo{year}{1997}),

\bibitem[{\citenamefont{Salman et~al.}(2011)\citenamefont{Salman, Smadella,
  MacFarlane, Patterson, Willmott, Chow, Hossain, Saadaoui, Wang, and
  Kiefl}}]{Salman2011}
\bibinfo{author}{\bibfnamefont{Z.}~\bibnamefont{Salman}},
  \bibinfo{author}{\bibfnamefont{M.}~\bibnamefont{Smadella}},
  \bibinfo{author}{\bibfnamefont{W.~A.} \bibnamefont{MacFarlane}},
  \bibinfo{author}{\bibfnamefont{B.~D.} \bibnamefont{Patterson}},
  \bibinfo{author}{\bibfnamefont{P.~R.} \bibnamefont{Willmott}},
  \bibinfo{author}{\bibfnamefont{K.~H.} \bibnamefont{Chow}},
  \bibinfo{author}{\bibfnamefont{M.~D.} \bibnamefont{Hossain}},
  \bibinfo{author}{\bibfnamefont{H.}~\bibnamefont{Saadaoui}},
  \bibinfo{author}{\bibfnamefont{D.}~\bibnamefont{Wang}}, \bibnamefont{and}
  \bibinfo{author}{\bibfnamefont{R.~F.} \bibnamefont{Kiefl}},
  \bibinfo{journal}{Phys. Rev. B} \textbf{\bibinfo{volume}{83}},
  \bibinfo{pages}{224112} (\bibinfo{year}{2011})

\bibitem[{\citenamefont{H{\"{u}}nnefeld
  et~al.}(2002)\citenamefont{H{\"{u}}nnefeld, Niem{\"{o}}ller, Schneider,
  R{\"{u}}tt, Rodewald, Fleig, and Shirane}}]{Hunnefeld2002}
\bibinfo{author}{\bibfnamefont{H.}~\bibnamefont{H{\"{u}}nnefeld}},
  \bibinfo{author}{\bibfnamefont{T.}~\bibnamefont{Niem{\"{o}}ller}},
  \bibinfo{author}{\bibfnamefont{J.}~\bibnamefont{Schneider}},
  \bibinfo{author}{\bibfnamefont{U.}~\bibnamefont{R{\"{u}}tt}},
  \bibinfo{author}{\bibfnamefont{S.}~\bibnamefont{Rodewald}},
  \bibinfo{author}{\bibfnamefont{J.}~\bibnamefont{Fleig}}, \bibnamefont{and}
  \bibinfo{author}{\bibfnamefont{G.}~\bibnamefont{Shirane}},
  \bibinfo{journal}{Phys. Rev. B} \textbf{\bibinfo{volume}{66}},
  \bibinfo{pages}{014113} (\bibinfo{year}{2002})

\bibitem[{\citenamefont{Loetzsch et~al.}(2010)\citenamefont{Loetzsch,
  Lu\"ubcke, Uschmann, Fo\"orster, Gro{\ss}e, Thuerk, Koettig, Schmidl, and
  Seidel}}]{Loetzsch2010}
\bibinfo{author}{\bibfnamefont{R.}~\bibnamefont{Loetzsch}},
  \bibinfo{author}{\bibfnamefont{A.}~\bibnamefont{Lu\"ubcke}},
  \bibinfo{author}{\bibfnamefont{I.}~\bibnamefont{Uschmann}},
  \bibinfo{author}{\bibfnamefont{E.}~\bibnamefont{Fo\"orster}},
  \bibinfo{author}{\bibfnamefont{V.}~\bibnamefont{Gro{\ss}e}},
  \bibinfo{author}{\bibfnamefont{M.}~\bibnamefont{Thuerk}},
  \bibinfo{author}{\bibfnamefont{T.}~\bibnamefont{Koettig}},
  \bibinfo{author}{\bibfnamefont{F.}~\bibnamefont{Schmidl}}, \bibnamefont{and}
  \bibinfo{author}{\bibfnamefont{P.}~\bibnamefont{Seidel}},
  \bibinfo{journal}{Appl. Phys. Lett.} \textbf{\bibinfo{volume}{96}},
  \bibinfo{pages}{071901} (\bibinfo{year}{2010})

\bibitem[{\citenamefont{Rehwald}(1970)}]{Rehwald1970}
\bibinfo{author}{\bibfnamefont{W.}~\bibnamefont{Rehwald}},
  \bibinfo{journal}{Solid State Commun.} \textbf{\bibinfo{volume}{8}},
  \bibinfo{pages}{1483} (\bibinfo{year}{1970})

\bibitem[{\citenamefont{Newnham}(2005)}]{Newnham2005}
\bibinfo{author}{\bibfnamefont{R.~E.} \bibnamefont{Newnham}},
  \emph{\bibinfo{title}{{Properties of Materials}}} (\bibinfo{publisher}{Oxford
  University Press}, \bibinfo{year}{2005})

\bibitem[{\citenamefont{Fossheim and Berre}(1972)}]{Fossheim1972}
\bibinfo{author}{\bibfnamefont{K.}~\bibnamefont{Fossheim}} \bibnamefont{and}
  \bibinfo{author}{\bibfnamefont{B.}~\bibnamefont{Berre}},
  \bibinfo{journal}{Phys. Rev. B} \textbf{\bibinfo{volume}{5}},
  \bibinfo{pages}{3292} (\bibinfo{year}{1972})

\bibitem[{\citenamefont{Fossum et~al.}(1984)\citenamefont{Fossum, Fossheim, and
  Scheel}}]{Fossum1984}
\bibinfo{author}{\bibfnamefont{J.}~\bibnamefont{Fossum}},
  \bibinfo{author}{\bibfnamefont{K.}~\bibnamefont{Fossheim}}, \bibnamefont{and}
  \bibinfo{author}{\bibfnamefont{H.~J.} \bibnamefont{Scheel}},
  \bibinfo{journal}{Solid State Commun.} \textbf{\bibinfo{volume}{51}},
  \bibinfo{pages}{839} (\bibinfo{year}{1984})

\bibitem[{\citenamefont{Rehwald}(1971)}]{Rehwald1971}
\bibinfo{author}{\bibfnamefont{W.}~\bibnamefont{Rehwald}},
  \bibinfo{journal}{Phys. der Kondens. Mater.} \textbf{\bibinfo{volume}{36}},
  \bibinfo{pages}{21} (\bibinfo{year}{1971})

\bibitem[{\citenamefont{Riste et~al.}(1971)\citenamefont{Riste, Samuelsen, and
  Otnes}}]{Riste1971}
\bibinfo{author}{\bibfnamefont{T.}~\bibnamefont{Riste}},
  \bibinfo{author}{\bibfnamefont{E.~J.} \bibnamefont{Samuelsen}},
  \bibnamefont{and} \bibinfo{author}{\bibfnamefont{K.}~\bibnamefont{Otnes}},
  \bibinfo{journal}{Solid State Commun.} \textbf{\bibinfo{volume}{17}},
  \bibinfo{pages}{1455} (\bibinfo{year}{1971}).

\bibitem[{\citenamefont{Darlington et~al.}(1975)\citenamefont{Darlington,
  Fitzgerald, and O'Connor}}]{Darlington1975}
\bibinfo{author}{\bibfnamefont{C.}~\bibnamefont{Darlington}},
  \bibinfo{author}{\bibfnamefont{W.}~\bibnamefont{Fitzgerald}},
  \bibnamefont{and} \bibinfo{author}{\bibfnamefont{D.}~\bibnamefont{O'Connor}},
  \bibinfo{journal}{Phys. Lett. A} \textbf{\bibinfo{volume}{54}},
  \bibinfo{pages}{35} (\bibinfo{year}{1975})

\bibitem[{\citenamefont{Fleury et~al.}(1968)\citenamefont{Fleury, Scott, and
  Worlock}}]{Fleury1968c}
\bibinfo{author}{\bibfnamefont{P.}~\bibnamefont{Fleury}},
  \bibinfo{author}{\bibfnamefont{J.}~\bibnamefont{Scott}}, \bibnamefont{and}
  \bibinfo{author}{\bibfnamefont{J.}~\bibnamefont{Worlock}},
  \bibinfo{journal}{Phys. Rev. Lett.} \textbf{\bibinfo{volume}{21}},
  \bibinfo{pages}{16} (\bibinfo{year}{1968})

\bibitem[{\citenamefont{M\"uller and Berlinger}(1971)}]{Muller1971}
\bibinfo{author}{\bibfnamefont{K.}~\bibnamefont{M\"uller}} \bibnamefont{and}
  \bibinfo{author}{\bibfnamefont{W.}~\bibnamefont{Berlinger}},
  \bibinfo{journal}{Phys. Rev. Lett.} \textbf{\bibinfo{volume}{26}},
  \bibinfo{pages}{13} (\bibinfo{year}{1971})

\bibitem[{\citenamefont{Bradler et~al.}(2009)\citenamefont{Bradler, Baum, and
  Riedle}}]{Bradler2009}
\bibinfo{author}{\bibfnamefont{M.}~\bibnamefont{Bradler}},
  \bibinfo{author}{\bibfnamefont{P.}~\bibnamefont{Baum}}, \bibnamefont{and}
  \bibinfo{author}{\bibfnamefont{E.}~\bibnamefont{Riedle}},
  \bibinfo{journal}{Appl. Phys. B Lasers Opt.} \textbf{\bibinfo{volume}{97}},
  \bibinfo{pages}{561} (\bibinfo{year}{2009})

\bibitem[{\citenamefont{Bojahr et~al.}(2013)\citenamefont{Bojahr, Herzog,
  Mitzscherling, Maerten, Schick, Goldshteyn, Leitenberger, Shayduk, Gaal, and
  Bargheer}}]{Bojahr2013}
\bibinfo{author}{\bibfnamefont{A.}~\bibnamefont{Bojahr}},
  \bibinfo{author}{\bibfnamefont{M.}~\bibnamefont{Herzog}},
  \bibinfo{author}{\bibfnamefont{S.}~\bibnamefont{Mitzscherling}},
  \bibinfo{author}{\bibfnamefont{L.}~\bibnamefont{Maerten}},
  \bibinfo{author}{\bibfnamefont{D.}~\bibnamefont{Schick}},
  \bibinfo{author}{\bibfnamefont{J.}~\bibnamefont{Goldshteyn}},
  \bibinfo{author}{\bibfnamefont{W.}~\bibnamefont{Leitenberger}},
  \bibinfo{author}{\bibfnamefont{R.}~\bibnamefont{Shayduk}},
  \bibinfo{author}{\bibfnamefont{P.}~\bibnamefont{Gaal}}, \bibnamefont{and}
  \bibinfo{author}{\bibfnamefont{M.}~\bibnamefont{Bargheer}},
  \bibinfo{journal}{Opt. Express} \textbf{\bibinfo{volume}{21}},
  \bibinfo{pages}{21188} (\bibinfo{year}{2013})

\bibitem[{\citenamefont{Bojahr et~al.}(2015)\citenamefont{Bojahr, Gohlke,
  Leitenberger, Pudell, Reinhardt, von Reppert, R\"{o}ssle, Sander, Gaal, and
  Bargheer}}]{Bojahr2015}
\bibinfo{author}{\bibfnamefont{A.}~\bibnamefont{Bojahr}},
  \bibinfo{author}{\bibfnamefont{M.}~\bibnamefont{Gohlke}},
  \bibinfo{author}{\bibfnamefont{W.}~\bibnamefont{Leitenberger}},
  \bibinfo{author}{\bibfnamefont{J.}~\bibnamefont{Pudell}},
  \bibinfo{author}{\bibfnamefont{M.}~\bibnamefont{Reinhardt}},
  \bibinfo{author}{\bibfnamefont{A.}~\bibnamefont{von Reppert}},
  \bibinfo{author}{\bibfnamefont{M.}~\bibnamefont{R\"{o}ssle}},
  \bibinfo{author}{\bibfnamefont{M.}~\bibnamefont{Sander}},
  \bibinfo{author}{\bibfnamefont{P.}~\bibnamefont{Gaal}}, \bibnamefont{and}
  \bibinfo{author}{\bibfnamefont{M.}~\bibnamefont{Bargheer}},
  \bibinfo{journal}{Phys. Rev. Lett.} \textbf{\bibinfo{volume}{115}},
  \bibinfo{pages}{195502} (\bibinfo{year}{2015}).

\bibitem[{\citenamefont{Ruello and Gusev}(2015)}]{Ruello2015}
\bibinfo{author}{\bibfnamefont{P.}~\bibnamefont{Ruello}} \bibnamefont{and}
  \bibinfo{author}{\bibfnamefont{V.~E.} \bibnamefont{Gusev}},
  \bibinfo{journal}{Ultrasonics} \textbf{\bibinfo{volume}{56}},
  \bibinfo{pages}{21} (\bibinfo{year}{2015})

\bibitem[{\citenamefont{Bojahr et~al.}(2012)\citenamefont{Bojahr, Herzog,
  Schick, Vrejoiu, and Bargheer}}]{Bojahr2012a}
\bibinfo{author}{\bibfnamefont{A.}~\bibnamefont{Bojahr}},
  \bibinfo{author}{\bibfnamefont{M.}~\bibnamefont{Herzog}},
  \bibinfo{author}{\bibfnamefont{D.}~\bibnamefont{Schick}},
  \bibinfo{author}{\bibfnamefont{I.}~\bibnamefont{Vrejoiu}}, \bibnamefont{and}
  \bibinfo{author}{\bibfnamefont{M.}~\bibnamefont{Bargheer}},
  \bibinfo{journal}{Phys. Rev. B} \textbf{\bibinfo{volume}{86}},
  \bibinfo{pages}{144306} (\bibinfo{year}{2012})

\bibitem[{\citenamefont{Cowley}(1968)}]{Cowley1968}
\bibinfo{author}{\bibfnamefont{R.~A.} \bibnamefont{Cowley}},
  \bibinfo{journal}{Reports Prog. Phys.} \textbf{\bibinfo{volume}{31}},
  \bibinfo{pages}{123} (\bibinfo{year}{1968})

\bibitem[{Note1()}]{Note1}
Note1, \bibinfo{note}{the phonon frequency is given by $\nu (q) = {2v_{\protect
  \text L}n(\lambda )\protect \qopname \relax o{cos}\theta }\protect \tmspace
  -\thinmuskip {.1667em}/\protect \tmspace -\thinmuskip {.1667em}_{\lambda }$
  with sound velocity $v_{\protect \text L}$, the wavelength dependent optical
  refractive index of STO $n(\lambda )$ taken from Ref.~\cite {Cardona1965},
  and the angle $\theta $ between surface normal and angle of incidence of the
  probe light.}

\bibitem[{\citenamefont{Maerten
  et~al.}(2015{\natexlab{b}})\citenamefont{Maerten, Bojahr, and
  Bargheer}}]{Maerten2015b}
\bibinfo{author}{\bibfnamefont{L.}~\bibnamefont{Maerten}},
  \bibinfo{author}{\bibfnamefont{A.}~\bibnamefont{Bojahr}}, \bibnamefont{and}
  \bibinfo{author}{\bibfnamefont{M.}~\bibnamefont{Bargheer}},
  \bibinfo{journal}{Ultrasonics} \textbf{\bibinfo{volume}{56}},
  \bibinfo{pages}{148} (\bibinfo{year}{2015}{\natexlab{b}}).

\bibitem[{\citenamefont{Koreeda and Saikan}(2011)}]{Koreeda2011}
\bibinfo{author}{\bibfnamefont{A.}~\bibnamefont{Koreeda}} \bibnamefont{and}
  \bibinfo{author}{\bibfnamefont{S.}~\bibnamefont{Saikan}},
  \bibinfo{journal}{Rev. Sci. Instrum.} \textbf{\bibinfo{volume}{82}},
  \bibinfo{pages}{126103} (\bibinfo{year}{2011})

\bibitem[{\citenamefont{Koreeda et~al.}(2006)\citenamefont{Koreeda, Nagano,
  Ohno, and Saikan}}]{Koreeda2006}
\bibinfo{author}{\bibfnamefont{A.}~\bibnamefont{Koreeda}},
  \bibinfo{author}{\bibfnamefont{T.}~\bibnamefont{Nagano}},
  \bibinfo{author}{\bibfnamefont{S.}~\bibnamefont{Ohno}}, \bibnamefont{and}
  \bibinfo{author}{\bibfnamefont{S.}~\bibnamefont{Saikan}},
  \bibinfo{journal}{Phys. Rev. B} \textbf{\bibinfo{volume}{73}},
  \bibinfo{pages}{024303} (\bibinfo{year}{2006})

\bibitem[{\citenamefont{Fano}(1961)}]{Fano1961}
\bibinfo{author}{\bibfnamefont{U.}~\bibnamefont{Fano}}, \bibinfo{journal}{Phys.
  Rev.} \textbf{\bibinfo{volume}{124}}, \bibinfo{pages}{1866}
  (\bibinfo{year}{1961})

\bibitem[{\citenamefont{Thomsen
  et~al.}(1986{\natexlab{a}})\citenamefont{Thomsen, Grahn, Maris, and
  Tauc}}]{Thomsen1986}
\bibinfo{author}{\bibfnamefont{C.}~\bibnamefont{Thomsen}},
  \bibinfo{author}{\bibfnamefont{H.}~\bibnamefont{Grahn}},
  \bibinfo{author}{\bibfnamefont{H.}~\bibnamefont{Maris}}, \bibnamefont{and}
  \bibinfo{author}{\bibfnamefont{J.}~\bibnamefont{Tauc}},
  \bibinfo{journal}{Phys. Rev. B} \textbf{\bibinfo{volume}{34}},
  \bibinfo{pages}{4129} (\bibinfo{year}{1986}{\natexlab{a}})

\bibitem[{\citenamefont{Thomsen
  et~al.}(1986{\natexlab{b}})\citenamefont{Thomsen, Grahn, Maris, and
  Tauc}}]{Thomsen1986a}
\bibinfo{author}{\bibfnamefont{C.}~\bibnamefont{Thomsen}},
  \bibinfo{author}{\bibfnamefont{H.}~\bibnamefont{Grahn}},
  \bibinfo{author}{\bibfnamefont{H.}~\bibnamefont{Maris}}, \bibnamefont{and}
  \bibinfo{author}{\bibfnamefont{J.}~\bibnamefont{Tauc}},
  \bibinfo{journal}{Opt. Commun.} \textbf{\bibinfo{volume}{60}},
  \bibinfo{pages}{55} (\bibinfo{year}{1986}{\natexlab{b}})

\bibitem[{\citenamefont{Binder and Knorr}(2001)}]{Binder2001}
\bibinfo{author}{\bibfnamefont{A.}~\bibnamefont{Binder}} \bibnamefont{and}
  \bibinfo{author}{\bibfnamefont{K.}~\bibnamefont{Knorr}},
  \bibinfo{journal}{Phys. Rev. B} \textbf{\bibinfo{volume}{63}},
  \bibinfo{pages}{094106} (\bibinfo{year}{2001})

\bibitem[{\citenamefont{Scott et~al.}(2012)\citenamefont{Scott, Salje, and
  Carpenter}}]{Scott2012a}
\bibinfo{author}{\bibfnamefont{J.~F.} \bibnamefont{Scott}},
  \bibinfo{author}{\bibfnamefont{E.~K.~H.} \bibnamefont{Salje}},
  \bibnamefont{and} \bibinfo{author}{\bibfnamefont{M.~A.}
  \bibnamefont{Carpenter}}, \bibinfo{journal}{Phys. Rev. Lett.}
  \textbf{\bibinfo{volume}{109}}, \bibinfo{pages}{187601}
  (\bibinfo{year}{2012})

\bibitem[{\citenamefont{Salje et~al.}(2013)\citenamefont{Salje, Aktas,
  Carpenter, Laguta, and Scott}}]{Salje2013}
\bibinfo{author}{\bibfnamefont{E.~K.~H.} \bibnamefont{Salje}},
  \bibinfo{author}{\bibfnamefont{O.}~\bibnamefont{Aktas}},
  \bibinfo{author}{\bibfnamefont{M.~A.} \bibnamefont{Carpenter}},
  \bibinfo{author}{\bibfnamefont{V.~V.} \bibnamefont{Laguta}},
  \bibnamefont{and} \bibinfo{author}{\bibfnamefont{J.~F.} \bibnamefont{Scott}},
  \bibinfo{journal}{Phys. Rev. Lett.} \textbf{\bibinfo{volume}{111}},
  \bibinfo{pages}{247603} (\bibinfo{year}{2013})

\bibitem[{\citenamefont{Salje et~al.}(2016)\citenamefont{Salje, Li, Stengel,
  Gumbsch, and Ding}}]{Salje2016}
\bibinfo{author}{\bibfnamefont{E.~K.~H.} \bibnamefont{Salje}},
  \bibinfo{author}{\bibfnamefont{S.}~\bibnamefont{Li}},
  \bibinfo{author}{\bibfnamefont{M.}~\bibnamefont{Stengel}},
  \bibinfo{author}{\bibfnamefont{P.}~\bibnamefont{Gumbsch}}, \bibnamefont{and}
  \bibinfo{author}{\bibfnamefont{X.}~\bibnamefont{Ding}},
  \bibinfo{journal}{Phys. Rev. B} \textbf{\bibinfo{volume}{94}},
  \bibinfo{pages}{024114} (\bibinfo{year}{2016})

\bibitem[{\citenamefont{Zykova-Timan and Salje}(2014)}]{Zykova-Timan2014}
\bibinfo{author}{\bibfnamefont{T.}~\bibnamefont{Zykova-Timan}}
  \bibnamefont{and} \bibinfo{author}{\bibfnamefont{E.~K.~H.}
  \bibnamefont{Salje}}, \bibinfo{journal}{Appl. Phys. Lett.}
  \textbf{\bibinfo{volume}{104}}, \bibinfo{pages}{082907}
  (\bibinfo{year}{2014})

\bibitem[{\citenamefont{Herzog et~al.}(2012)\citenamefont{Herzog, Bojahr,
  Goldshteyn, Leitenberger, Vrejoiu, Khakhulin, Wulff, Shayduk, Gaal, and
  Bargheer}}]{Herzog2012a}
\bibinfo{author}{\bibfnamefont{M.}~\bibnamefont{Herzog}},
  \bibinfo{author}{\bibfnamefont{A.}~\bibnamefont{Bojahr}},
  \bibinfo{author}{\bibfnamefont{J.}~\bibnamefont{Goldshteyn}},
  \bibinfo{author}{\bibfnamefont{W.}~\bibnamefont{Leitenberger}},
  \bibinfo{author}{\bibfnamefont{I.}~\bibnamefont{Vrejoiu}},
  \bibinfo{author}{\bibfnamefont{D.}~\bibnamefont{Khakhulin}},
  \bibinfo{author}{\bibfnamefont{M.}~\bibnamefont{Wulff}},
  \bibinfo{author}{\bibfnamefont{R.}~\bibnamefont{Shayduk}},
  \bibinfo{author}{\bibfnamefont{P.}~\bibnamefont{Gaal}}, \bibnamefont{and}
  \bibinfo{author}{\bibfnamefont{M.}~\bibnamefont{Bargheer}},
  \bibinfo{journal}{Appl. Phys. Lett.} \textbf{\bibinfo{volume}{100}},
  \bibinfo{pages}{094101} (\bibinfo{year}{2012})

\bibitem[{\citenamefont{Nagakubo et~al.}(2012)\citenamefont{Nagakubo, Yamamoto,
  Tanigaki, Ogi, Nakamura, and Hirao}}]{Nagakubo2012}
\bibinfo{author}{\bibfnamefont{A.}~\bibnamefont{Nagakubo}},
  \bibinfo{author}{\bibfnamefont{A.}~\bibnamefont{Yamamoto}},
  \bibinfo{author}{\bibfnamefont{K.}~\bibnamefont{Tanigaki}},
  \bibinfo{author}{\bibfnamefont{H.}~\bibnamefont{Ogi}},
  \bibinfo{author}{\bibfnamefont{N.}~\bibnamefont{Nakamura}}, \bibnamefont{and}
  \bibinfo{author}{\bibfnamefont{M.}~\bibnamefont{Hirao}},
  \bibinfo{journal}{Jpn. J. Appl. Phys.} \textbf{\bibinfo{volume}{51}},
  \bibinfo{pages}{07GA09} (\bibinfo{year}{2012})

\bibitem[{\citenamefont{Shayduk et~al.}(2013)\citenamefont{Shayduk, Herzog,
  Bojahr, Schick, Gaal, Leitenberger, Navirian, Sander, Goldshteyn, Vrejoiu
  et~al.}}]{Shayduk2013}
\bibinfo{author}{\bibfnamefont{R.}~\bibnamefont{Shayduk}},
  \bibinfo{author}{\bibfnamefont{M.}~\bibnamefont{Herzog}},
  \bibinfo{author}{\bibfnamefont{A.}~\bibnamefont{Bojahr}},
  \bibinfo{author}{\bibfnamefont{D.}~\bibnamefont{Schick}},
  \bibinfo{author}{\bibfnamefont{P.}~\bibnamefont{Gaal}},
  \bibinfo{author}{\bibfnamefont{W.}~\bibnamefont{Leitenberger}},
  \bibinfo{author}{\bibfnamefont{H.}~\bibnamefont{Navirian}},
  \bibinfo{author}{\bibfnamefont{M.}~\bibnamefont{Sander}},
  \bibinfo{author}{\bibfnamefont{J.}~\bibnamefont{Goldshteyn}},
  \bibinfo{author}{\bibfnamefont{I.}~\bibnamefont{Vrejoiu}},
  \bibnamefont{et~al.}, \bibinfo{journal}{Phys. Rev. B}
  \textbf{\bibinfo{volume}{87}}, \bibinfo{pages}{184301}
  (\bibinfo{year}{2013})

\bibitem[{\citenamefont{Akhiezer}(1939)}]{Akhiezer1939}
\bibinfo{author}{\bibfnamefont{A.}~\bibnamefont{Akhiezer}},
  \bibinfo{journal}{J. Phys. USSR} p. \bibinfo{pages}{277}
  (\bibinfo{year}{1939}).

\bibitem[{Note2()}]{Note2}
Note2, \bibinfo{note}{in the linear inset we shifted the maxima of $\Gamma (T)$
  to $T_{\protect \text {a}}\protect \tmspace -\thinmuskip {.1667em}=\protect
  \tmspace -\thinmuskip {.1667em}105$\protect \tmspace +\thinmuskip {.1667em}K.
  We have obtained the ``exact'' transition temperatures by the fit using the
  function $A\cdot |T - T_{\protect \text {a}}|^{-\eta }$ which yields the
  slope $\eta $ of the linear fit together with a fit constant $A$, which
  represents the offset of the abscissa. The values of $\eta $ ($\eta '$) and
  $T_{\protect \text {a}}$ are given in Tab.~\ref {tab:ta-damping}.}

\bibitem[{\citenamefont{Guennou et~al.}(2010)\citenamefont{Guennou, Bouvier,
  Kreisel, and Machon}}]{Guennou2010}
\bibinfo{author}{\bibfnamefont{M.}~\bibnamefont{Guennou}},
  \bibinfo{author}{\bibfnamefont{P.}~\bibnamefont{Bouvier}},
  \bibinfo{author}{\bibfnamefont{J.}~\bibnamefont{Kreisel}}, \bibnamefont{and}
  \bibinfo{author}{\bibfnamefont{D.}~\bibnamefont{Machon}},
  \bibinfo{journal}{Phys. Rev. B} \textbf{\bibinfo{volume}{81}},
  \bibinfo{pages}{054115} (\bibinfo{year}{2010})

\bibitem[{\citenamefont{B{\"{a}}uerle and Rehwald}(1978)}]{Bauerle1978a}
\bibinfo{author}{\bibfnamefont{D.}~\bibnamefont{B{\"{a}}uerle}}
  \bibnamefont{and} \bibinfo{author}{\bibfnamefont{W.}~\bibnamefont{Rehwald}},
  \bibinfo{journal}{Solid State Commun.} \textbf{\bibinfo{volume}{27}},
  \bibinfo{pages}{1343} (\bibinfo{year}{1978}),

\bibitem[{\citenamefont{von Waldkirch et~al.}(1972)\citenamefont{von Waldkirch,
  M{\"{u}}ller, Berlinger, and Thomas}}]{VonWaldkirch1972}
\bibinfo{author}{\bibfnamefont{T.}~\bibnamefont{von Waldkirch}},
  \bibinfo{author}{\bibfnamefont{K.~A.} \bibnamefont{M{\"{u}}ller}},
  \bibinfo{author}{\bibfnamefont{W.}~\bibnamefont{Berlinger}},
  \bibnamefont{and} \bibinfo{author}{\bibfnamefont{H.}~\bibnamefont{Thomas}},
  \bibinfo{journal}{Phys. Rev. Lett.} \textbf{\bibinfo{volume}{28}},
  \bibinfo{pages}{503} (\bibinfo{year}{1972})

\bibitem[{\citenamefont{Cardona}(1965)}]{Cardona1965}
\bibinfo{author}{\bibfnamefont{M.}~\bibnamefont{Cardona}},
  \bibinfo{journal}{Phys. Rev.} \textbf{\bibinfo{volume}{140}},
  \bibinfo{pages}{A651} (\bibinfo{year}{1965})

\end{thebibliography}
\end{document}